\documentclass[proof]{WileyASNA-v1}

\articletype{Article Type}%

\received{30 July 2021}
\revised{DD MON 2021}
\accepted{DD MON 2021}

\begin{document}

\title{High-resolution observations of low-luminosity peaked spectrum sources}

\author[1,2,3]{Jordan D. Collier}

\authormark{Collier 2021}

\address[1]{The Inter-University Institute for Data Intensive Astronomy (IDIA), Department of Astronomy, University of Cape Town, Private Bag X3, Rondebosch, 7701, South Africa}

\address[2]{School of Science, Western Sydney University, Locked Bag 1797, Penrith, NSW 2751, Australia}

\address[3]{CSIRO Astronomy and Space Science, PO Box 1130, Bentley, WA, 6102, Australia}



\abstract{Although radio observations have been historically seen as less valuable than optical observations, today's broadband radio spectra of peaked spectrum sources reveal detailed physics from within the inner region of the galaxy, on spatial scales beyond what an optical telescope can resolve. Peaked radio spectra are thought to be evolving into large scale radio galaxies, although an over-abundance of the most compact sources reveals that a significant fraction are confined within their host galaxies. Furthermore, at the lowest luminosities, these sources are largely unknown, and may reveal the small scale precursors of FR-I galaxies. Here I summarise the previous work exploring the properties of low luminosity peaked radio sources, and the future work that extends on this within even deeper radio observations of well studied fields.}

\keywords{active galaxies, radio continuum, galaxy evolution, image processing}

\jnlcitation{\cname{%
\author{J. D. Collier}} (\cyear{2021}), 
\ctitle{High-resolution observations of low-luminosity PS sources}, \cjournal{Astron. Nachr}, \cvol{2021;XX:Y--Z}.}


\maketitle


\section{Introduction}

Radio observations have typically historically been seen as less valuable compared to observations from optical and IR telescopes, with some citing quotations such as ``there's nothing as useless as a radio source''. Even from an interferometer, radio observations typically have a spatial resolution of a few to tens of arcsec, and a characteristic power-law spectral index that many argue doesn't tell you much about the physical properties of a source (e.g. a galaxy), certainly not compared to an optical observation with a spatial resolution of $\lesssim$1 arcsec. 

While that may be true, the current era of broadband radio observations, with large bandwidths over single instruments, and even larger multi-wavelength bandwidths over multiple radio telescopes, allows for much detail to be determined from radio observations alone, and for the most compact radio galaxies ($\sim$5-20\% of the radio sky), gives you detailed physics from within the inner region of the galaxy.

For the population of peaked spectrum (PS) radio sources, such as Gigahertz-Peaked Spectrum (GPS), and the related class of Compact Steep-Spectrum (CSS) sources, the entirety of the radio emission comes from within a region far smaller than what an optical telescope is able to resolve, and in fact, far smaller than the host galaxy. 

\begin{figure}
\includegraphics[width=0.5\textwidth]{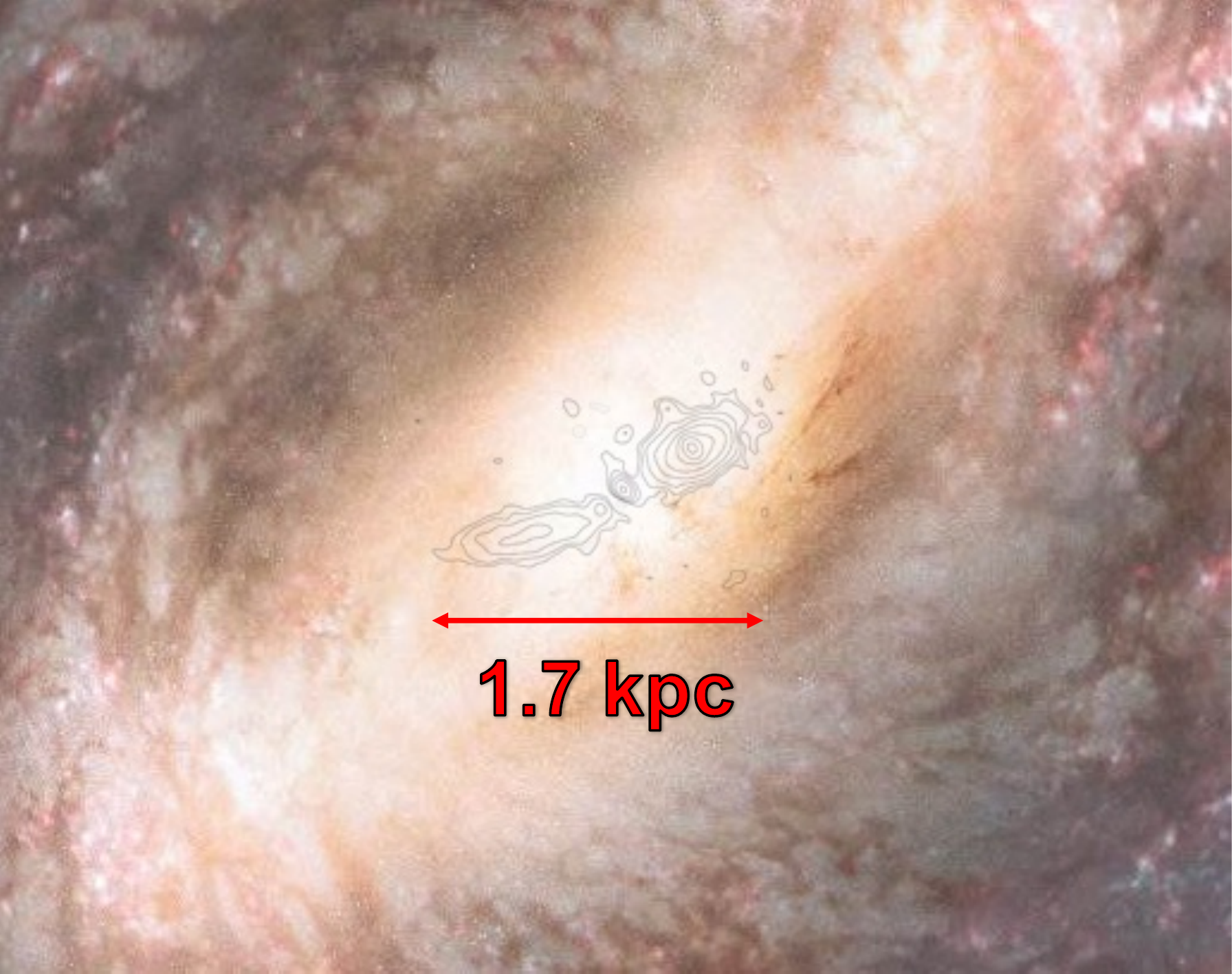}
\caption{A typical CSS source, F00183-7111, with contours showing the entirety of the radio emission, compared to the scale of the inner bulge of the Milky Way.}
\label{size-comparison}
\end{figure}

To put this into perspective, a typical CSS source, F00183-7111 \citep{2012MNRAS.422.1453N}, 1.7 kpc in size, is shown by the contours in Figure~\ref{size-comparison}, compared to the inner bulge of the Milky Way galaxy, $\sim$30 kpc in size. Given that all of the radio emission, observed over a broad radio spectrum, is entirely encapsulated by the region shown, any physics derived from the radio emission, such as the absorption mechanism, electron ageing, magnetic field strengths, and jet power, constrains the physical properties of the inner region of the galaxy. Therefore, radio observations of peaked radio spectra alone are valuable for probing the formation and evolution of galaxies across cosmic time, including the role of the inner Supermassive Black Hole (SMBH) and its effect on the galaxy. Adding optical and IR to radio data within a multi-wavelength study even further enables an understanding of PS sources, in particular, the redshifts and luminosities from optical spectra.

\subsection{Peaked Radio Spectra}

As they evolve, PS sources are thought to grown into CSS sources, which are thought to be growing into large scale FR-I/II galaxies. However, the over-abundance of the most compact radio galaxies \citep{An12} has led to alternative theories, suggesting that they could be: 1) confined galaxies within a dense medium; 2) transient or intermittent radio galaxies, each of which includes observed examples. While the evolutionary scenario is likely a function of the abundance of material accreting onto the SMBH, and each scenario may represent a fraction of the entire population, the prevalence of each scenario is not well known.

At the lowest luminosities, PS sources are largely unknown \citep{2015arXiv151201851S}, and it has also been proposed that the low-luminosity population includes the missing precursors to FR-I galaxies \citep{2015arXiv151009061K,2015MNRAS.448..252T}. However, very few examples are known, due to the lack of high-resolution observations within deep radio surveys that include low-frequency coverage. 

The turnover of a peaked radio spectrum is an absorption effect that is most commonly modelled by Synchrotron Self-Absorption (SSA), with the radio flux density given by

\begin{equation}
S_{\nu} = a\left(\frac{\nu}{\nu_m}\right)^{-(\beta-1)/2}\left(\frac{1-e^{-\tau}}{\tau}\right),
\label{SSA_eqn}
\end{equation}

\noindent where $a$ is a flux density normalisation parameter, $\nu_m$ is the turnover frequency, $\beta$ is the power-law index of the electron energy distribution, and $\tau$ is the optical depth given by $(\nu/\nu_m)^{-(\beta+4)/2}$.
%
%
%
%
%
%
Although SSA is likely present in the majority of cases, there is an ongoing debate around the prevalence of Free-Free Absorption (FFA).

Another feature that may be present in the radio spectrum of PS and CSS sources is a spectral break, an electron ageing effect in which higher energy electrons expend their energy more quickly, resulting in fewer emitting electrons with increasing frequency, causing a steepening in the high frequency spectrum. In the continuous injection model \citep{1962SvA.....6..317K}, the jet continually ejects fresh electrons into the lobe of the radio galaxy, causing a sharp break in the spectrum, which steepens to $\alpha-0.5$ above the break frequency $\nu_{\rm br}$.
%
%
In the exponential break model \citep{1973A&A....26..423J}, the jet switches off and the spectrum exponentially steepens after the break frequency, given by $e^{-\nu/\nu_{\rm br}}$.

\section{Previous Work}

In our previous work \citep{2018MNRAS.477..578C}, we compiled a low luminosity sample of PS and CSS sources and observed them with Very Long Baseline Interferometry (VLBI). This was selected from amongst the sample presented in \citet{2015arXiv151107929C} and \citet{2016PhDT.......266C}, which contains 71 PS and CSS sources with $L_{\rm 1.4 GHz} = 10^{21-27}$ W\,Hz$^{-1}$, selected from two of the deepest fields, the Chandra Deep Field South (CDFS) and European Large Area ISO Survey South 1 (ELAIS-S1). The previous work from \citet{2018MNRAS.477..578C} is summarised below.

In this work, we observed eight PS and CSS sources with VLBI, and detected six of them, which revealed linear sizes between 0.07--1.8 kpc, and redshifts between $0.26 < z < 1.1$. Variability was ruled out in most cases, mainly based on close frequency measurements from different epochs, although one source was proposed to be a variable quasar, for which a peaked radio spectrum was temporarily adopted during a flaring phase. This interpretation was strengthened by the presence of a flat spectrum derived from the simultaneous 5.5 and 9 GHz observations, and led to discarding the source form our analysis. All sources were resolved on at least one axis, and one was a classical double radio source, although we did not resolve the source enough to determine whether it resembled an FR-I or FR-II radio galaxy.

We modelled the radio spectra using frequencies between 72 MHz and 34 GHz, using the Levenberg-Marquardt algorithm, and comparing the model selection with the Bayesian information criterion (BIC). If no curvature was seen, a power-law model was fit, otherwise both SSA and the inhomogeneous FFA model from \citet{bic97} were fit and evaluated according to their BIC values. If the high frequency spectra were observed to deviate from a power law, both spectral break models were fit, and the one with the lowest BIC value was selected. For four of the six sources, one was described equally well by a broken power law and FFA, one marginally favoured SSA over FFA (both with an exponential break), one was described equally well by SSA and FFA, and one was described equally well by SSA with a spectral break, and FFA. The last two included one source fit by a power law model, and another suggested to be a variable quasar. All alternative models to FFA but one also required a spectral break to accurately model the radio spectrum.

We derived statistical model ages of 1.7--5.1 kyr based on the jet growth of the sample from \citet{An12}. We compared these to the spectral ages derived from the SSA radio spectra, assuming equipartitional magnetic field strengths, which were 0.7--2.7 kyr. Both age estimates were generally within agreement with each other, and were consistent with the youth hypothesis on the nature of PS and CSS sources. One source revealed an exponential break with a turnoff time of 0.3 kyr. This may be an intermittent source, that may switch back on if new material is accreted onto the SMBH. We compared the position of the sources in the turnover frequency - linear size diagram to those from \citet{2014MNRAS.438..463O}, and found consistency between the two, suggesting that low luminosity PS and CSS sources also co-evolve in linear size and turnover frequency, although two were marginal outliers on the diagram, given that they had upper limits on their turnover frequencies.

\section{Future Work}

Since the number of peaked radio sources increases with decreasing turnover frequency, selecting from low-frequency samples increases the sample size. Selecting samples from deep, high-resolution broadband observations that include low frequencies will enable the low-luminosity population to be uncovered, and increase the number of known peaked radio sources manyfold. This would also include compact Megahertz-Peaked Spectrum (MPS) sources, which are known to include high-redshift radio galaxies \citep[HzRGs;][]{2015MNRAS.450.1477C,2016MNRAS.463.3260C,2016MNRAS.459.2455C}, whose low frequency turnovers suggest large sizes, but whose observed compact sizes suggest they are highly redshifted GPS sources. Selecting such sources from the Upgraded Giant Metrewave Radio Telescope (uGMRT), with a turnover frequency at ~400 MHz, allows us to trace GPS sources peaking from 1.2-1.6 GHz between redshifts $2 < z < 3$. 

\subsection{Deep ELAIS-N1 sample}
\label{EN1}

To extend on the work presented in \citet{2018MNRAS.477..578C}, a new even deeper sample is being compiled that has much more complementary depth and resolution at low frequencies. This is a sample from the ELAIS North 1 (ELAIS-N1) region, and includes 150 MHz data from the Low Frequency Array (LOFAR) with an RMS of $\sim$20\,$\mu$Jy/beam \citep{2021A&A...648A...2S}, as well as deep images from the (upgraded) GMRT, over a deep field, and a wide field. The deep field includes uGMRT observations of the inner $\sim$10\% of the ELAIS-N1 field at about 400 and 600 MHz, down to an RMS of $\sim$20 and $\sim$12\,$\mu$Jy/beam, respectively. The wide field includes 325 and 610 MHz observations, down to an RMS of $\sim$250 and $\sim$40\,$\mu$Jy/beam, respectively. All these data have a comparable spatial resolution at $\lesssim$10 arcsec, and therefore avoid issues with blended components between frequencies. 

In addition to the ELAIS-N1 data listed above, this field includes 1400 MHz Very Large Array (VLA) data from the FIRST survey \citep{Becker1995} at $\sim$100 $\mu$Jy/beam, deeper and higher resolution VLA data down to an RMS of 87 $\mu$Jy/beam \citep{2011ApJ...733...69B}, and even deeper VLA data to which we have access, over $\sim$0.1 deg$^2$ down to an RMS of $\sim$1\,$\mu$Jy/beam. Lastly, we also have Dominion Radio Astrophysical Observatory (DRAO) data within this field down to an RMS of $\sim$40 $\mu$Jy/beam \citep{2010ApJ...714.1689G}, although with $\sim$1 arcmin resolution.

\begin{figure}
\includegraphics[width=0.5\textwidth]{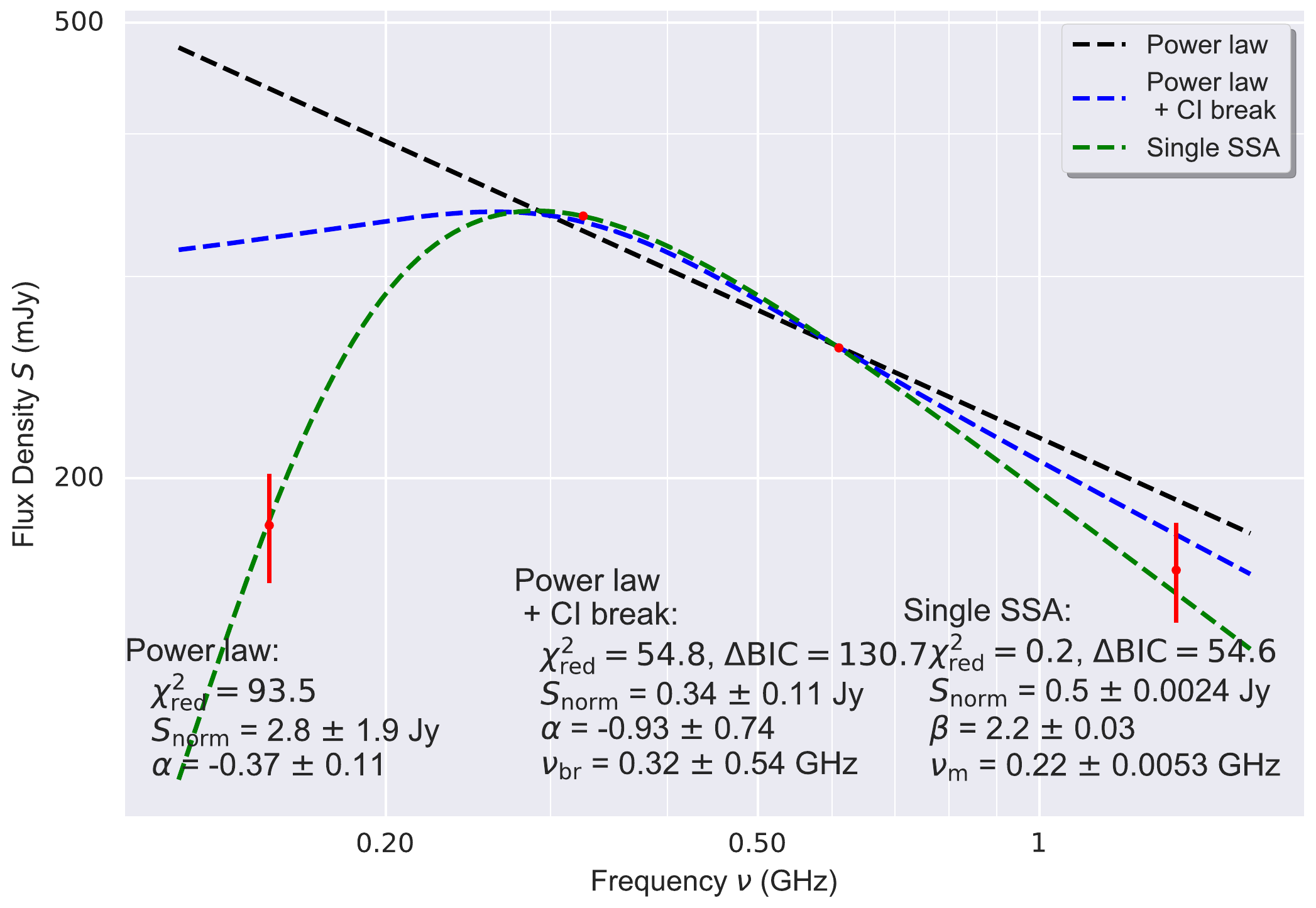}
\caption{An example spectrum from our ELAIS-N1 sample, with a power-law, broken power-law and SSA models fit for comparison. The information at the bottom of the figure shows the fitted parameters and uncertainties, the $\chi_{\rm red}^2$, and the difference in BIC values, to inform model selection.}
\label{SED-example}
\end{figure}

As an initial exploration of these data (using FIRST for the 1400 MHz data), we performed a nearest neighbour cross-match within 3 arcsec, and fit radio spectra to explore the spectral properties of the sample, and identify any peaked spectra. This included fitting broken power laws and an SSA model, and comparing the models based on their $\chi_{\rm red}^2$ and BIC values, to inform model selection. A typical example of a peaked source with the SSA model preferred is shown in Figure~\ref{SED-example}.

\begin{figure}
\includegraphics[width=0.5\textwidth]{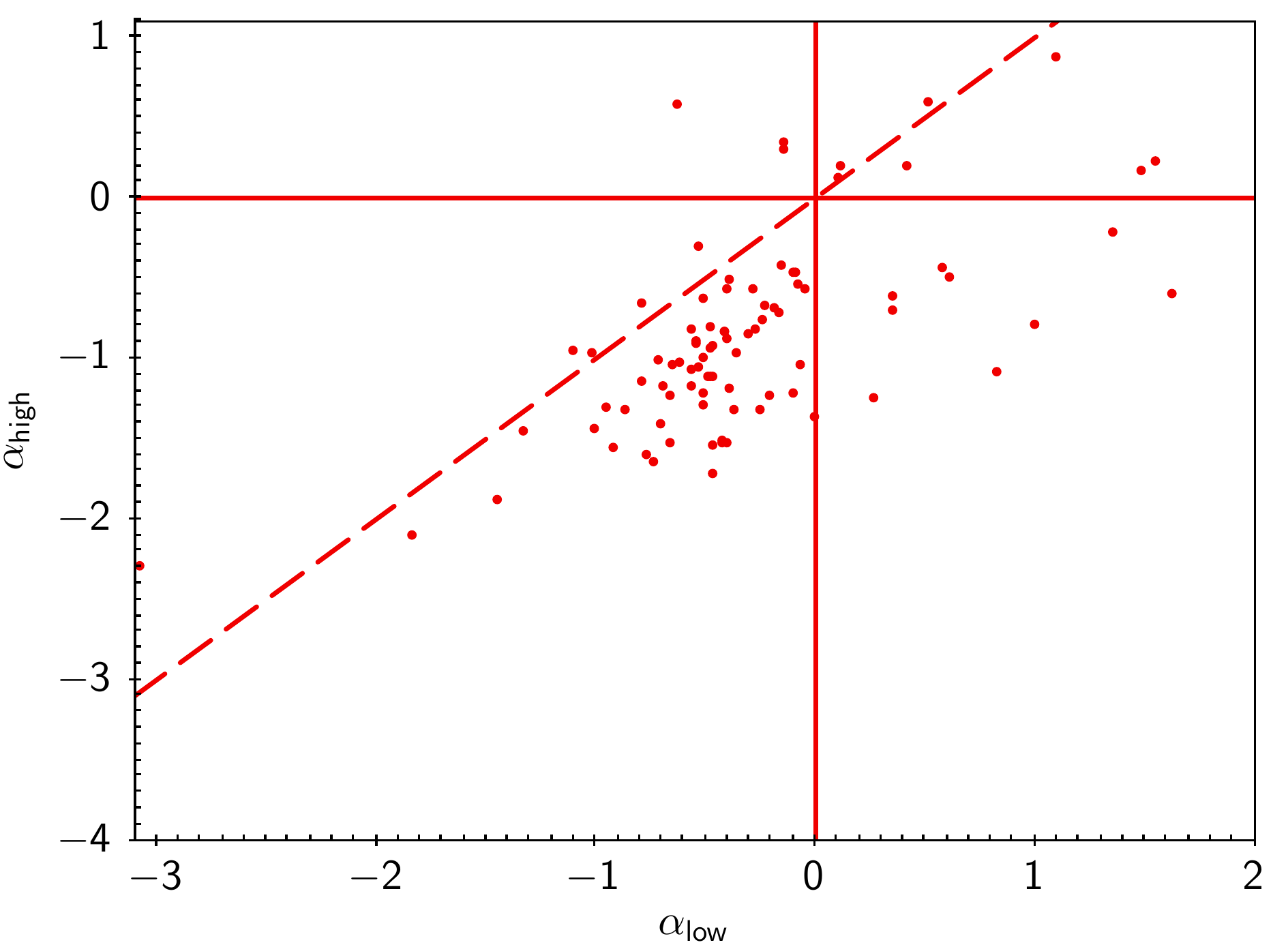}
\includegraphics[width=0.5\textwidth]{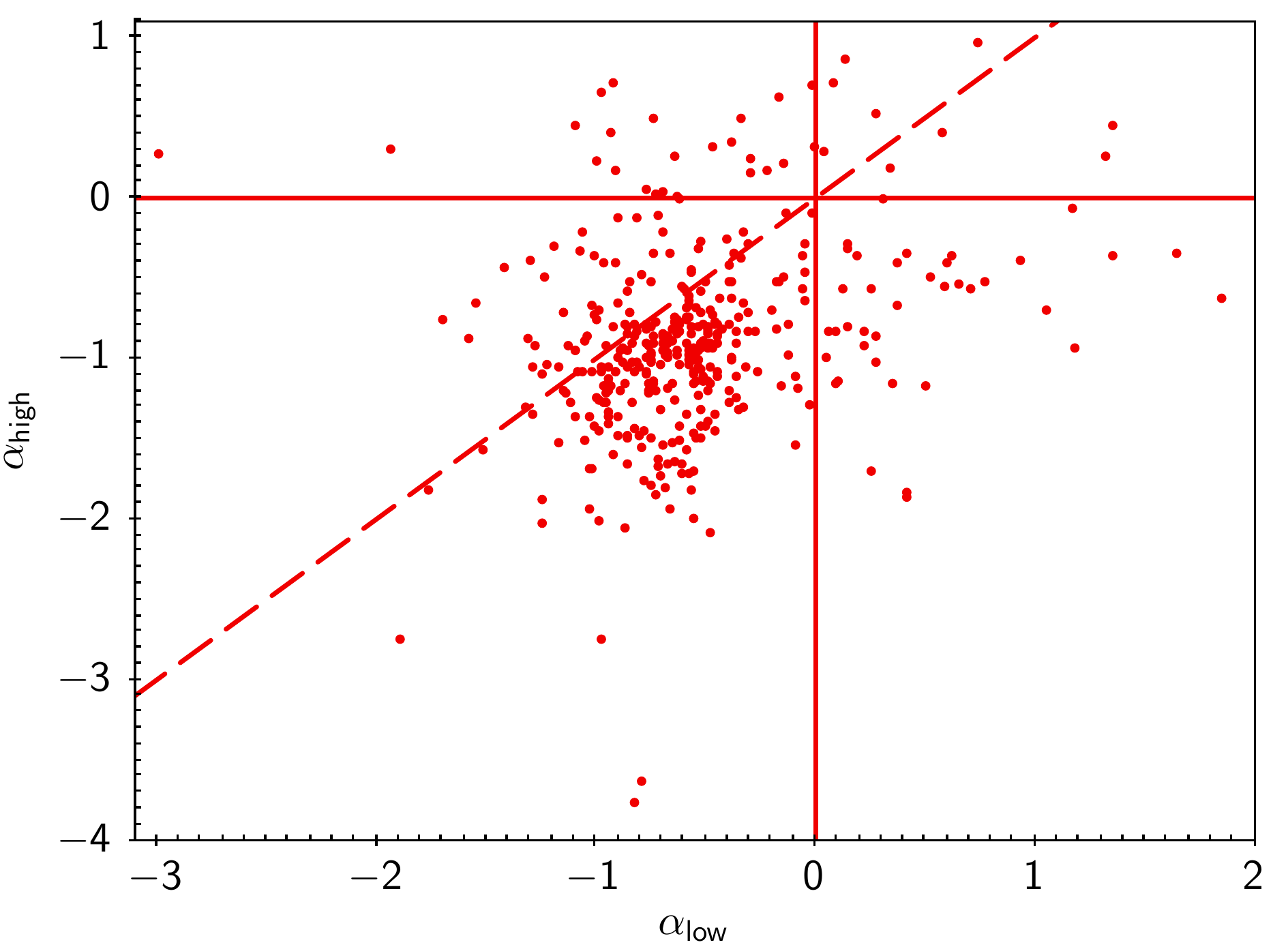}
\caption{The radio colour-colour diagram from our preliminary ELAIS-N1 deep (top panel) and wide (bottom panel) sample, compiled by cross-matching the data discussed in Section~\ref{EN1} within 3 arcsec, and separately fitting a power law to the low- and high-frequency spectra.}
\label{colour-colour}
\end{figure}

Additionally, we explored the radio colour-colour properties of the sample by fitting a power law to the low frequency spectra, between 150 and 325/400 MHz, and the high frequency spectra, between 610 and 1400 MHz. The resulting radio colour-colour diagram is shown in Figure~\ref{colour-colour}. This shows similar properties between the deep and wide samples, with the majority of sources exhibiting a spectral index of $\alpha_{\rm low} \sim -0.8$, steeping a little to $\alpha_{\rm high} \sim -1.0$. Additionally, the sample contains a number of peaked, convex and inverted spectra. We will extend this sample by including more data and by also fitting a low and high frequency power law to data that only contain measurements at three frequencies.

With this sample, one could possibly compile the deepest and lowest luminosity sample to date, select PS/CSS/MPS sources, and characterise the properties of the sample, using the myriad of multi-wavelength data available within this field, determining their redshifts, luminosities, and star formation rates. One could examine how their properties compare to the well-known brighter samples, using evolutionary tracers such as their spectral ages, linear size and turnover frequency to understand whether the sample represents a continuous population of young and evolving radio galaxies, and the prevalence of other evolutionary scenarios within the sample. Following up with VLBI observation over a sub-sample of the most interesting objects, including candidate FR-I precursors, and HzRGs selected from MPS sources, would enable the linear sizes and morphologies of the most compact and low-luminosity sources to be determined, to test whether they exhibit a compact morphology analogous to an FR-I source. VLBI observations may also enable monitoring of jet growth to determine dynamical ages.

\bibliography{Wiley-ASNA}

\section*{Author Biography}

\begin{biography}{
\includegraphics[width=0.2\textwidth]{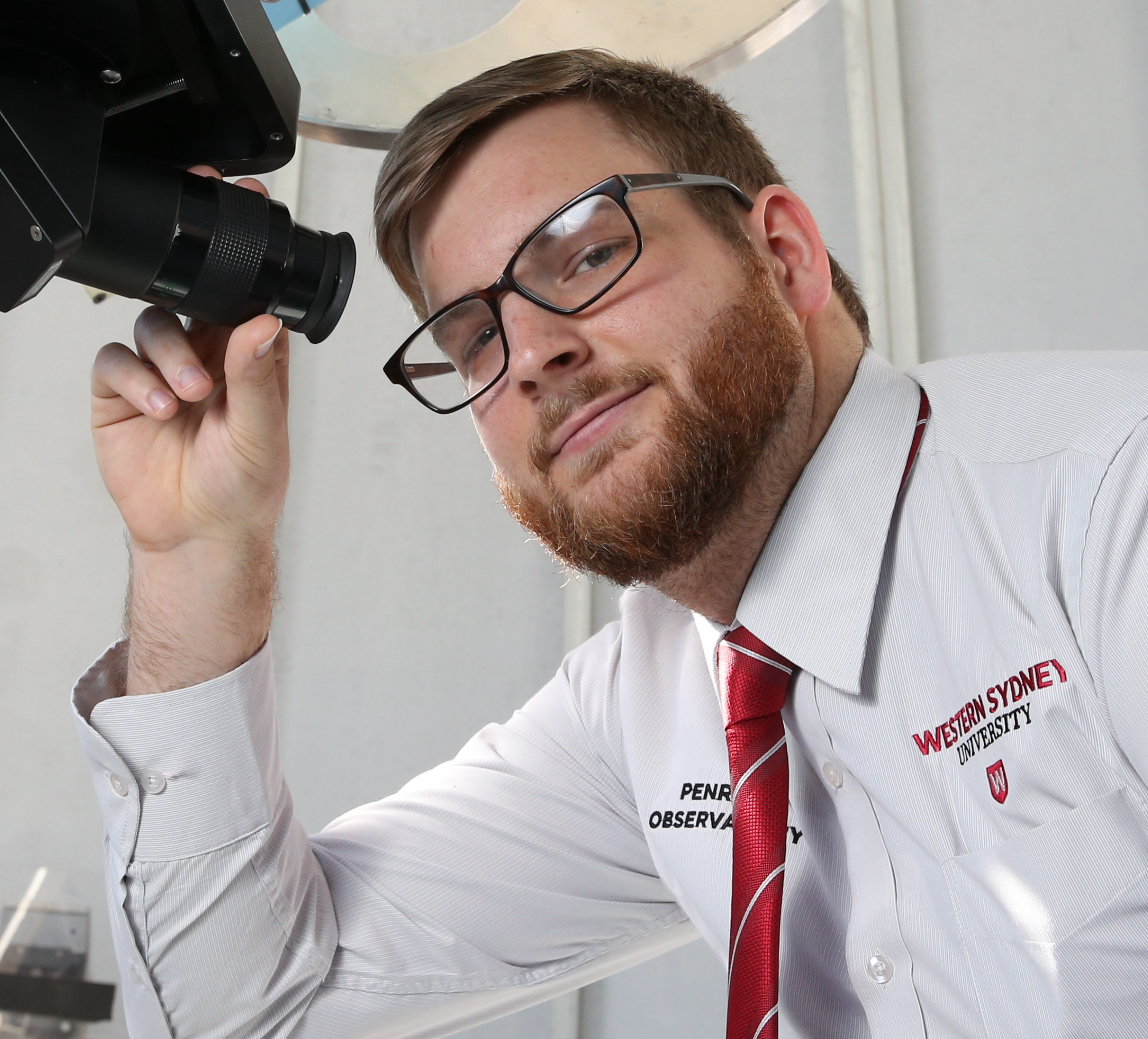}
}
{\textbf{Dr Jordan D. Collier.} \href{http://www.ilifu.ac.za/}{ilifu} Support Astronomer at \href{https://idia.ac.za}{IDIA}, with a PhD in Astronomy and Astrophysics, and an interest in supercomputing, large and deep radio surveys, radio astronomy data processing techniques, galaxy evolution within the radio continuum, and young and distant radio galaxies.}

\end{biography}

\end{document}